\documentclass[%
reprint,
superscriptaddress,
aps,
]{revtex4-1}

\usepackage{graphicx}
\usepackage{hyperref}
\hypersetup{colorlinks=true, linkcolor=blue, urlcolor=dblue, citecolor=blue}
\usepackage{amsmath} \usepackage{amssymb} \usepackage{upgreek}
\usepackage{accents} \usepackage{mathrsfs} \usepackage{setspace}
\usepackage{color}
\usepackage{braket}
\definecolor{dgreen}{RGB}{00, 120, 00} \definecolor{dblue}{RGB}{00, 00, 180}
\definecolor{lgreen}{RGB}{46, 139, 87} 
\definecolor{orange}{RGB}{255, 165, 00}

\usepackage{ulem}

\newcommand{\buf}[1]{\textcolor{blue}{#1}} 
\newcommand{\bs}[1]{\boldsymbol{#1}}

\newcommand{\ut}[1]{\undertilde{#1}}

\begin{document}

%
%

\title{Chirality Detection through Vortex Bound States in ($d+id'$)-Wave Superconductor}

\author{Soma Yoshida}
\affiliation{Department of Applied Physics, Nagoya University, Nagoya 464-8603, Japan }
\author{Yukio Tanaka}
\affiliation{Department of Applied Physics, Nagoya University, Nagoya 464-8603, Japan }
\author{Alexander A. Golubov}
\affiliation{Faculty of Science and Technology and MESA+ Institute for Nanotechnology, University of Twente, 
7500 AE Enschede, The Netherlands}
\author{Shu-Ichiro Suzuki}
\email[Email:]{s.suzuki-1@utwente.nl}
\affiliation{Faculty of Science and Technology and MESA+ Institute for Nanotechnology, University of Twente, 
7500 AE Enschede, The Netherlands}

\date{\today}

\begin{abstract}

	We present a method for detecting the chirality $\chi$ of a ($d_{zx}+i \chi
d_{yz}$)-wave superconductor through the analysis of the local density
of states (LDOS) at the vortex core. Employing the quasiclassical
Eilenberger theory, we examine the LDOS in a semi-infinite
superconductor with a quantum vortex 
penetrating the surface perpendicularly.  
We show that $\mathrm{sgn}[\chi]$ changes completely  
the LDOS at the core-surface intersection. Remarkably, the difference
between LDOS for the $\chi = 1$ and $\chi = -1$ states becomes more
prominent when the surface is dirtier, meaning that 
one does not need to pay close
attention to the surface quality of the sample. 
The difference between these two states arises from the symmetry of
the subdominant Cooper pairs induced at the core-surface intersection:
whether the subdominant $s$-wave Cooper pairs are present or not.
Due to the unique nature of this phenomenon in the ($d_{zx}+i \chi
d_{yz}$)-wave superconductor, one can potentially demonstrate the
realization of the ($d_{zx}+i \chi d_{yz}$)-wave superconductivity and determine its chirality by, for instance, through scanning tunnel spectroscopy experiments.
\end{abstract}

\pacs{pacs}

\thispagestyle{empty}

\maketitle


\textit{Introduction.}--Additional spontaneous symmetry breaking associated with orbital angular momenta of Cooper pairs stands as a distinctive characteristic of chiral superconductors (SCs). 
The chiral superconductivity is realized by Cooper pairs with finite orbital
angular momenta \cite{Leggett_RMP_1975,  Rice_Sigrist_1995,  kita_JPSJ_98, Matsumoto_JPSJ_99, Furusaki_PRB_01, Stone_PRB_04, Braunecker_PRL_05,
Nagato_JPSJ_11, Sauls_PRB_11, Bakurskiy_PRB_14, Kallin_PRB_14, Tada_PRL_15, Kallin_RPP_2016,  Higashitani_JPSJ_20, Suzuki_PRR_22, SIS_PRB_23}. 
At the superconducting transition temperature $T_c$, the angular momenta of all Cooper pairs align in the same direction: upwards ($\chi=+1$) or downwards ($\chi=-1$) where the chirality $\chi$ characterize the degenerated two states.  This selection leads to additional spontaneous symmetry breaking. 
Detecting the chirality is essential in a comprehensive understanding of the chiral nature of superconductivity.
However, in a long history of superconductivity and superfluidity, $\chi$ has only been detected in one single experiment of the superfluid $^3$He A-phase \cite{Ikegami_Science_2013, Ikegami_JPSJ_2015},
where the intrinsic Magnus force on impurities flowing through $^3$He was employed to identify $\chi$ \cite{Salmelin_PRL_1989, Salmelin_PRB_1990, Ikegami_Science_2013, Ikegami_JPSJ_2015}. 
On the contrary, in \textit{superconductivity}, experimental observation of $\chi$ has never been succeeded \cite{Moler_PRB_05, Moler_PRB_07}. 

Signs of chiral superconductivity have been observed by, for instance, muon-spin relaxation/rotation ($\mu$SR) experiments \cite{Luke_Nature_1998, Grinenko_NP_2020, Grinenko_NP_2021}.
The emergence of internal magnetic fields along a certain axis has
been detected below $T_c$, suggesting the orbital motion of Cooper
pairs. 
However, the $\mu$SR technique cannot identify $\chi$ as the muon depolarisation rate is independent of the sign of the local field. 
To demonstrate the realization of chiral superconductivity, one needs a method that can directly access $\chi$, the signal arising from the orbital angular momentum by the Cooper pairs. 
In this Letter, we will propose a novel approach for local detection of $\chi$ utilizing the
interaction between two different bound states: the surface Andreev bound
state (ABS)\cite{Buchholtz_PRB_1981, Hara_Nagai_1986, Hu_PRL_1994,
Tanaka_PRL_1995} and the Caroli-de~Gennes-Matricon (CdGM) mode
\cite{caroli_PL_1964, kramer_ZP_1974,
hess_PRL_1989,Ullah_PRB_1990, Gygi_PRB_1991, Schopohl_PRB_1995, Ichioka_PRB_1996}. Such an interaction drastically modifies
the quasiparticle local density of states (LDOS) that is commonly
measured by, for example, scanning tunneling spectroscopy (STS) \cite{hess_PRL_1989}. 

\begin{figure}[b]
	\centering
	\includegraphics[width=0.8\columnwidth]{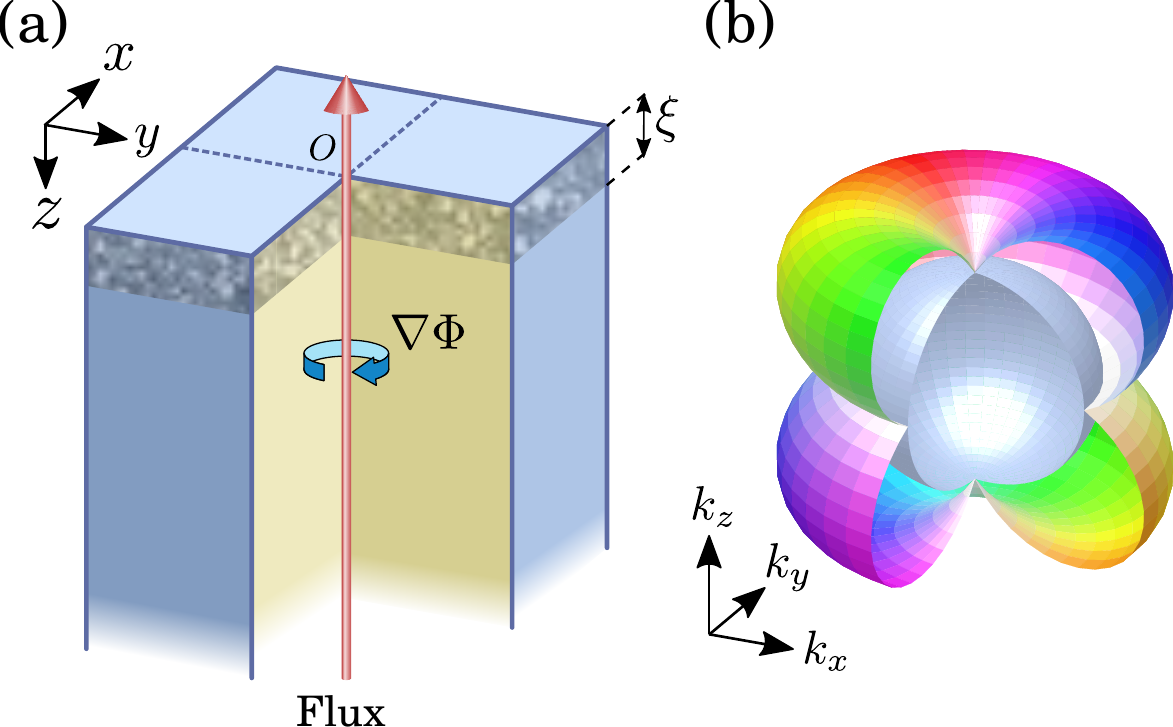}
	\caption{(a) Semi-infinite 3D SC occupying $z \geq 0$ and quantum
	vortex at $x=y=0$. The surface is disordered at  $0 \leq z \leq \xi$.
	(B) Gap amplitudes of the ($d_{xz}+id_{yz}$)-wave SCs. The
	color represents the phase winding of the pair potential. 
	The inner sphere indicates the Fermi sphere.} 
	\label{fig:model}
\end{figure}

We consider a semi-infinite three-dimensional (3D) chiral SC with a
quantum flux penetrating the surface perpendicularly
[Fig.~\ref{fig:model}(a)]. 
We assume the $(d_{zx}+id_{yz})$-wave pairing, which could be realized in
URu$_2$Si$_2$~\cite{Kasahara_PRL_2007, Kasahara_NJP_2009, Schemm_PRB_2015},
Sr$_2$RuO$_4$~\cite{pustogow_Nature_2019, Suh_PRR_2020,
Grinenko_NatPhys_2021, grinenko2021unsplit, Ikegaya_PRR_21, Suzuki_PRR_22}, and
LaPt$_3$P~\cite{Biswas_NC_2021}. 
Employing the quasiclassical Eilenberger formalism, we examine the
quasiparticle LDOS under the self-consistent
pair potential. 
The calculated results show that $\chi$ completely changes the LDOS at
the core-surface intersection. Remarkably, the difference in the LDOS
for the $\chi= \pm 1$ states becomes the more prominent when the
surface is the dirtier, meaning that one does not need to pay close
attention to the surface quality of the sample. 

The different behaviors depending on $\chi$ can be coherently interpreted in terms of the subdominant Cooper pairs which are induced by the local symmetry breaking. In the antiparallel configuration (APC) ($\chi=-1$), the subdominant Cooper pairs have the isotropic $s$-wave pairing symmetry, which is robust against surface roughness thanks to the isotropic nature\cite{Kato_JPSJ_2001, Kato_JPSJ_2002, Tanuma_PRL_2009, Suzuki_PRB_2015}. In the parallel configuration (PC) ($\chi=1$), on the other hand, the leading component has the ($d_{x^2-y^2}+i\chi d_{xy}$)-wave symmetry which is fragile against roughness.

\textit{Model and formulation.}--
We consider a single vortex perpendicular to the surface of a 3D SC
occupying $z\ge0$ [Fig.~\ref{fig:model}(a)]. The vortex is located at
$\rho=0$.
Throughout this Letter, we use cylindrical coordinates $\bs{r}=(x,
y, z) = (\rho \cos \Phi, \rho \sin \Phi,z)$ and assume the rotational
symmetry around the vortex core. 
We employ the quasiclassical Eilenberger formalism to analyze the LDOS
and the symmetry of Cooper pairs. The quasiclassical Green's function
(GF) $\check{g}=\check{g}(\bs{r},\hat{\bs{k}},i\omega_n)$ 
obeys the Eilenberger equation \cite{Eilenberger_ZP_1968}, 
\begin{align}
  &\bs{v}_F \cdot \bs{\nabla} \check{g}
  +[\tau_3 \omega_n + \check{H},\check{g}]=0,
	\quad
	\check{g}^2= \check{1}, \label{eq:eilenberger}
	\\
  &\check{g}=\left( \begin{array}{rr}
          g  &      f  \\
      \ut{f} & -\ut{g}
      \end{array}
  \right),
	\hspace{4mm}
  \check{H}=\left( \begin{array}{cc}
	  \upxi & \upeta \\
	  \ut{\upeta} & -\upxi \\
    \end{array} \right), \\
	& \upxi  =          \Gamma \langle g \rangle, \hspace{4mm}
	\upeta = \Delta(\hat{\bs{k}},\bs{r}) + \Gamma \langle f \rangle, 
		\label{eq:eff}
\end{align}
where $\bs{v}_F = (k_F/m) \hat{\bs{k}}$ is the Fermi velocity, 
$\hat{\bs{k}} = (k_x, k_y, k_z) = 
(\sin \theta_k \cos \phi_k,  
 \sin \theta_k \sin \phi_k, 
 \cos \theta_k)$, 
 and Matsubara frequency $\omega_n=(2n+1)\pi T$ with $T$ and $n$ being the temperature and an integer number. 
The type-II limit ($\lambda_L \gg \xi$) is considered. 
The surface roughness is taken into account through $\Gamma\langle g\rangle$ and $\Gamma\langle f\rangle$
within the self-consistent Born approximation, 
where the impurity scattering rate $\Gamma$ is finite
only near the surface $z \leq \xi$ \footnote{The mean free path $\ell$  
and $\Gamma$ are related by the relation: 
$\ell / \xi_0=\pi T_c / \Gamma \approx 1.78 \Delta_0/ \Gamma$
, where $\xi_0=\hbar v_F/2\pi k_B T_c$ is the coherent length.
} and $\braket{\cdots}=\int\cdots d\Omega/4\pi$ denotes
the angle average on the Fermi sphere.
For an arbitrary function $X$, $\ut{X}$
	is defined as
$\ut{X}(\hat{\bs{k}},\bs{r},i\omega_n)=X^*(-\hat{\bs{k}},\bs{r},i\omega_n)$.
The Eilenberger equation is solved by the so-called Riccati parameterization \cite{Schopohl_PRB_1995, Eschrig_PRB_00,
Eschrig_PRB_09}.
In this Letter, we refer to the anomalous GF $f$ as the pair amplitude. 
Throughout this Letter, we use the units $\hbar=c=k_B=1$.

We consider the ($d_{zx}+i\chi d_{yz}$)-wave [($d+i\chi d'$)-wave] SC 
which is a typical model for spin-singlet chiral SCs
\cite{Kasahara_PRL_2007, Kasahara_NJP_2009,
Schemm_PRB_2015, pustogow_Nature_2019, Suh_PRR_2020,
Grinenko_NatPhys_2021, grinenko2021unsplit, Ikegaya_PRR_21, Biswas_NC_2021, Suzuki_PRR_22}. 
In the presence of a vortex,
the pair potential for the ($d+i\chi d'$)-wave SC is given by
\cite{Sauls_NJP_2009, Eschrig_NJP_2009, Suh_PRR_2020, Patric_PRB_23}
\begin{align}
	\Delta(\hat{\bs{k}}, \bs{r})
	&=\Delta_{ \chi}(\rho,z)\psi_{ \chi}(\hat{\bs{k}})e^{iM\Phi}\nonumber\\
	&+\Delta_{-\chi}(\rho,z)\psi_{-\chi}(\hat{\bs{k}})e^{i(M+2\chi)\Phi},\label{eq:pp}\\
	\psi_{\pm1}(\hat{\bs{k}})&=2k_z(k_x\pm ik_y),
\end{align} 
where $\Delta_{\chi}$ is the dominant component, $\Delta_{-\chi}$ is the subdominant component induced with opposite chirality to the principal component, and $M$ is the vorticity. The subdominant
component vanishes at $\rho \gg \xi$ and $z \gg \xi$.  
Hereafter, we denote $\Delta_{\pm1}$ and $\psi_{\pm1}$ by $\Delta_{\pm}$ and $\psi_{\pm}$, respectively.
The function $\psi_\pm$ represents the momentum dependence of the
pair potential. As schematically shown in Fig.~\ref{fig:model}(b), the 
$(d+id')$-wave SC has a horizontal node at $k_z=0$ and point nodes at
the north and south poles. 
In the homogeneous limit, the chiral SC spontaneously chooses 
	either of $\chi=1$ or $-1$ state from the two degenerated states 
	that triggers the time reversal symmetry breaking. In this letter, therefore, we consider two distinct
	configurations: an APC (i.e., $\chi=-M$)
	and a PC (i.e., $\chi=M$).

The pair potential $\Delta_{\pm}(\bs{r})$ are determined self-consistently by the 
gap equation, 
\begin{align}
	\Delta_{\pm}(\bs{r})
	&=2\lambda N_0 \pi k_B T\sum_{n=0}^{N_{\mathrm{c}}}
	\braket{V_{\pm}(\hat{\bs{k}})
	f(\hat{\bs{k}},\bs{r},i\omega_n)},\label{eq:gap}
	\\
	\lambda
	&=N_0^{-1} \left(\sum_{n=0}^{N_{\mathrm{c}}}
	\frac{1}{n+1/2}+\ln\frac{T}{T_c}\right)^{-1}\nonumber
\end{align}
where 
$N_0$ is the density of states (DOS) per spin at the Fermi
level in the normal
state and $N_c$ is the positive integer satisfying
$N_c<\Omega_{c}/2\pi T$ with 
$\Omega_{c}$ being the cut-off
frequency. 
The attractive potential $V_{\pm}$ is represented as $V_{\pm}(\bs{k})=15\psi_{\pm}(\hat{\bs{k}})/8$.
Solving Eqs.~\eqref{eq:eilenberger}-\eqref{eq:eff} and \eqref{eq:gap} numerically, 
we obtain the quasiclassical GF self-consistently.

The LDOS $N(\bs{r},E)$ is given by
\begin{align}
	N(\bs{r},E)
	={N_0}
	\langle \mathrm{Re}[g(\hat{\bs{k}},\bs{r},E+i\delta)] \rangle, 
\end{align}
where $E$ is the energy. Throughout this letter, the parameters are set to:
$\Omega_c=40T_c$, $T=0.2T_c$, and $\delta = 0.04\Delta_0$
with $\Delta_0=\Delta_\chi(\rho\rightarrow\infty,z\rightarrow\infty)$ being the gap amplitude in the uniform system.


\begin{figure}[tbp]
	\centering
	\includegraphics[width=0.92\columnwidth]{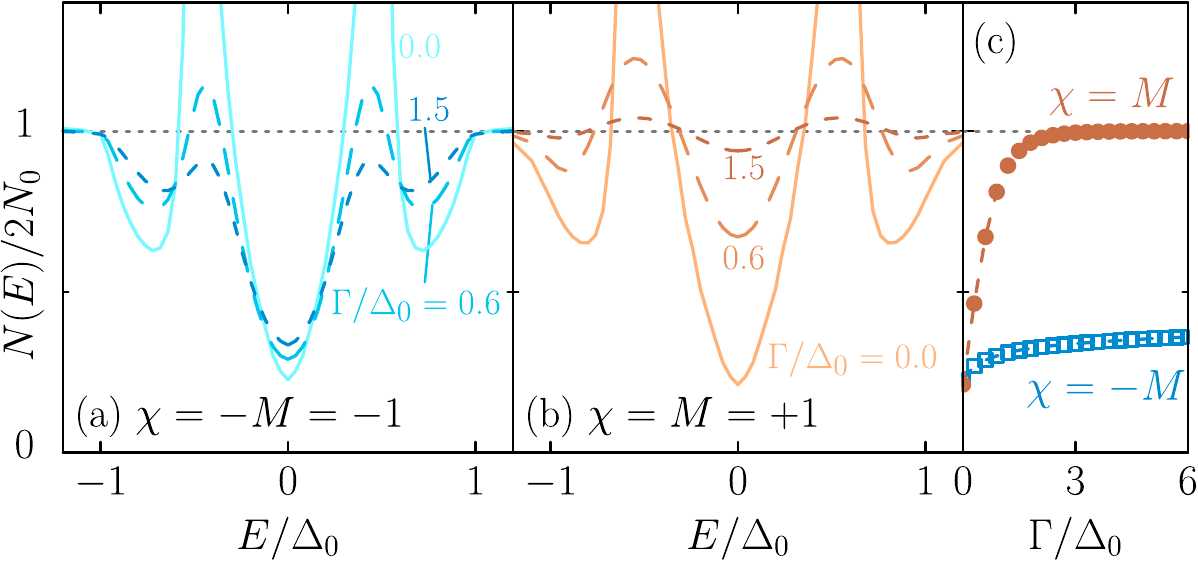}
	\caption{
		Effect of surface roughness on the LDOS at $\rho=z=0$. 
	The LDOS at (a) $\chi=M$ and (b) $\chi=-M$ are shown for $\Gamma/\Delta_0=0.0, 0.6$, and $1.5$.
	(c) Impurity scattering rate $\Gamma$ dependence of the zero energy LDOS at $\rho=z=0$.  
	The zero energy dip at $\chi=-M$ is robust against the non-magnetic
	impurity scattering, while that is fragile at $\chi=M$. 
	}
	\label{fig:ldos-Gammadep}
\end{figure}

\textit{Local density of states.}--We first show the LDOS at the
core-surface intersection (i.e., $\rho=z=0$) in
Fig.~\ref{fig:ldos-Gammadep}, where (a) $\chi=-M$ and (b) $\chi=+M$
and $\Gamma/\Delta_0=0.0$, $0.6$, and $1.5$.
Figure~\ref{fig:ldos-Gammadep} shows that $\chi$ completely changes
the LDOS at the vortex core. When $\chi=-M$
[Fig.~\ref{fig:ldos-Gammadep}(a)], the LDOS show a dip structure at
$E=0$ regardless of the surface roughness. When $\chi=+M$
[Fig.~\ref{fig:ldos-Gammadep}(b)], on the contrary, the zero-energy
dip (ZED) is significantly smeared by the surface roughness.  When
$\Gamma/\Delta_0 = 1.5$ which corresponds to $\ell/\xi_0
\approx 1.0$ with $\ell$ being the mean free path, the LDOS becomes $N(E) \sim N_0$, meaning
that the superconductivity is killed near the surface. 
Therefore, by
measuring $N(\bs{r})|_{E=0}$ at the top surface of a $(d+id')$-wave
SC with vortices, it is possible to illustrate a chiral-domain map
$\chi(\bs{r})$, where $N|_{E=0}$ has a deep (shallow) dip at a
positive (negative) chirality domain.

\begin{figure}[t]
	\centering
	\includegraphics[width=0.96\columnwidth]{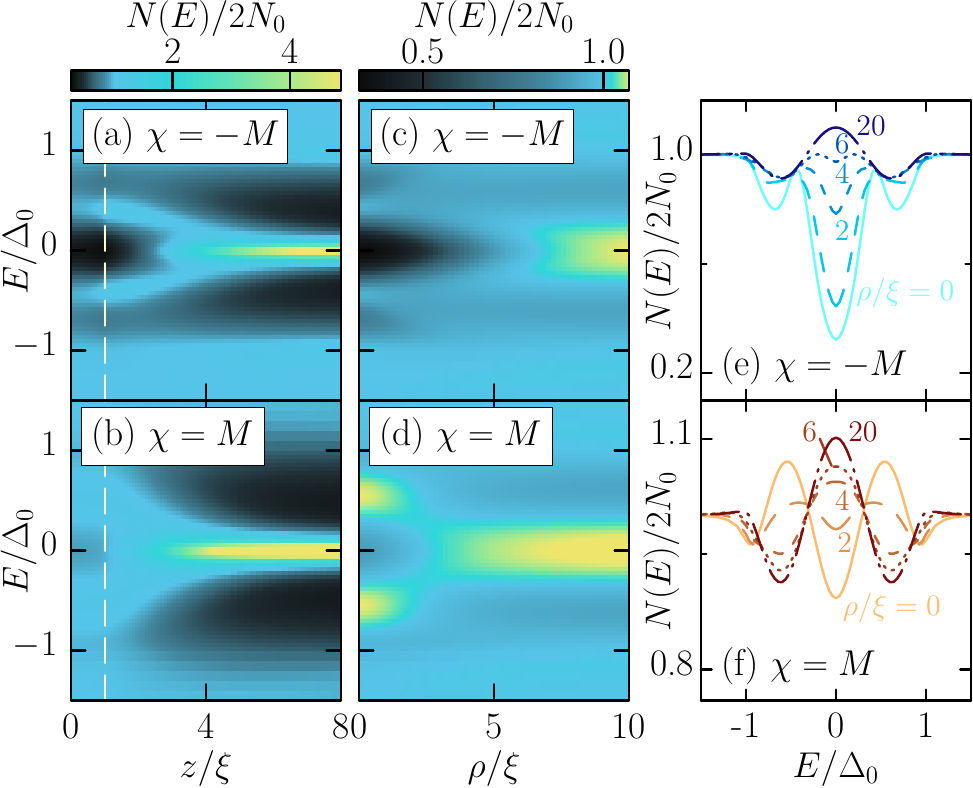}
	\caption{Local density of states \textbf{(a,b)} along the vortex core and \textbf{(c-f)} at the top surface. The surface roughness is set to $\Gamma=1.2\Delta_0$. The other parameters are set to the same values as used in Fig.~\ref{fig:ldos-Gammadep}. The dashed lines in (a,b) indicate the border between the disordered and clean regions ($z=\xi$).
	} 
        \label{fig:ldos-map}
\end{figure}

The $\Gamma$-dependence of $N|_{E=0}$ are shown in
Fig.~\ref{fig:ldos-Gammadep}(c). For $\chi=+M$, $N|_{E=0}$ is
significantly affected by the roughness and approaches its normal
value $N_0$ even with weak roughness $\Gamma < \Delta_0$.
On the contrary, $N|_{E=0}$ for $\chi=-M$ nearly depends on $\Gamma$.
Even under the significant roughness (i.e., $\Gamma/\Delta_0=6$),
the LDOS remains $N|_{E=0} \approx 0.3N_0$. Remarkably, the difference
in $N|_{E=0}$ in each $\chi$ region increases with increasing
roughness $\Gamma$. Namely, in this novel approach, one does not need to
pay attention to the surface quality of the sample. 

The dip structure in the LDOS appears not just as a result of the
level repulsion of two ZES: surface ABS and CdGM modes.  The LDOS
along the vortex core ($\rho=0$) are shown in
Fig.~\ref{fig:ldos-map}(a,b), where $\Gamma/\Delta_0 = 1.2$.  Deep
inside the SC, the LDOS shows the ZEP (i.e., CdGM modes) regardless of
$\chi$. In the clean limit, the CdGM modes interact with the surface
ABSs near surface \footnote{The $(d_{zx}+id_{yz})$-wave SC hosts
flat-band ABSs at the surfaces in the $z$ direction. The condition for
the flat-band ABS is given by $\Delta(k_\perp, k_\parallel) =
-\Delta(-k_\perp, k_\parallel)$. }. 
The peak appears no longer at $E=0$ but two split peaks appear around $|E|
\sim 0.5 \Delta_0$ at $\rho=z=0$ [Figs.~\ref{fig:ldos-Gammadep}(a,b)
and \ref{fig:ldos-map}(a,b)].
In general, surface roughness kills \textit{only} the flat-band ABS
\cite{tanaka_PRB_2007, tanaka_JPSJ_2012, Suzuki_PRR_22} because surface roughness is a disordered localized near the surface.
The disappearance of the flat-band ABS, however, does not recover the ZEP of CdGM modes
as shown in Figs~\ref{fig:ldos-Gammadep}(a,b): The surface roughness
changes only the height of the finite-energy peaks but not their energy. 
When $\chi = -M$, furthermore, the depth of the ZED is
barely affected by the roughness. 
The interplay between the phase winding and the surface reflection induces the unique effective superconductivity at the intersection, which keeps the ZED even under the strong roughness (We will discuss the details later).

The LDOS at the top surface ($z=0$), measurable quantity by the STS technique \cite{hess_PRL_1989}, is shown in Fig.~\ref{fig:ldos-map}(c-f),  where $\Gamma=1.2\Delta_0$. 
Far from the vortex, the zero-energy peak (ZEP) appears regardless of the chirality due to the flat-band ABS \cite{Hu_PRL_1994, sato_PRB_2011, tanaka_JPSJ_2012, Shingo_PRB_2015, Yoshida_PRR_22, Suzuki_PRR_22}, 
where the height of the ZEP is not as high as its height in the clean limit. 
With approaching the vortex core, $N(E)|_{E=0}$ changes from a peak to a dip structure. The dip area depends on $\chi$ as well as the dip depth. 
When $\chi=-M$, the vortex bound state with a dip structure extends to  $\rho \sim 5\xi_0$ [Fig.~\ref{fig:ldos-map}(a)],  which is several times larger than the typical vortex states \cite{Tanuma_PRL_2009}.
When $\chi=M$, the ZED is confined within a narrower area ($\rho < 2\xi_0$) than the $\chi=-M$ case.

\begin{figure}[tb]
	\centering
	\includegraphics[width=0.97\columnwidth]{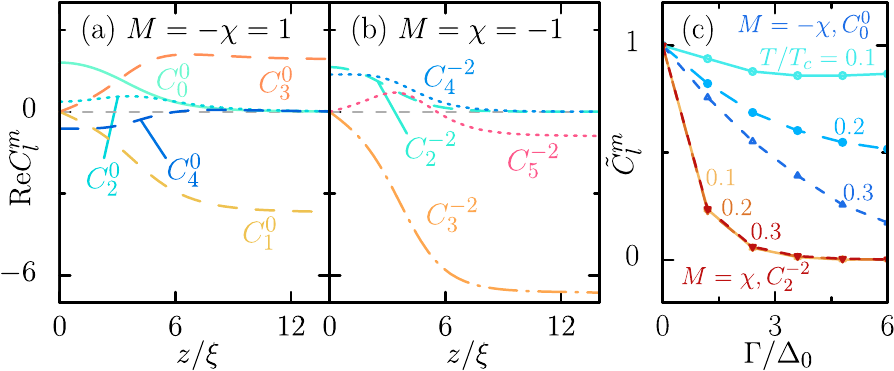}
	\caption{Pair amplitudes at the vortex core. 
	The Matsubara frequency and the surface roughness are set to
	$\omega_0$ and $\Gamma=0$. 
	The results for the APC and PC are shown in
	\textbf{(a)} and \textbf{(b)}, respectively. \textbf{(c)} Roughness
	dependence of the pair amplitudes. The pair amplitudes are
	normalized to their values for $\Gamma=0$; $\tilde{C}_l^m =
	\mathrm{Re}[C_l^m/C_l^m|_{\Gamma=0}]$. The $C_2^{-2}$ component
	in the PC decays rapidly with increasing the roughness
	$\Gamma$, whereas the $C_0^{0}$ component in the APC 
	survives under the
	roughness. The pair function is more robust at low temperatures. 
	}
	\label{fig:pa}
\end{figure}

\textit{Subdominant Cooper pairs.}--The robustness of the
zero-energy dip (ZED) is determined by whether the $s$-wave pairs are
induced at the core-surface intersection as subdominant pairs.
Additional local symmetry breaking can induce different types of Cooper
pairs from the principal component [i.e., $(d+i \chi d')$-wave] \cite{tanaka_PRB_2007}. In
the APC, the ZED is robust because it is supported by
$s$-wave subdominant pairs, whereas the ZED in the PC is
easily smeared out because the $s$-wave pair is absent.  The
difference in the LDOS with respect to $\chi$ can be coherently
understood in terms of the pairing symmetry of the subdominant
component around the vortex core.

We here explain the additional local symmetry breaking and 
the pairing symmetry of the induced subdominant pairs. 
The momentum dependence of $f$ can be expanded in terms of 
the spherical harmonics $Y_l^m$, 
\begin{align}
	& f_n (\bs{r}, \hat{\bs{k}}) 
	= \sum_{l m} C_{l,n}^{m}(\bs{r}) Y_l^{m}(\hat{\bs{k}}), 
	\label{ha_ex}
\end{align}
where 
{$f_n (\bs{r}, \hat{\bs{k}}) 
= f   (\bs{r}, \hat{\bs{k}}, i\omega_n) $
 and } we focus on
$\Phi=0$ because of the rotational symmetry\cite{yokoyama_PRB_2008, tanaka_JPSJ_2012}. 
Equation \eqref{ha_ex} contains odd-parity Cooper pairs with odd $l$. These pairs must be
antisymmetric with respect to the frequency (i.e., odd-frequency Cooper pairs\cite{Berezinskii}) 
so that the pair function satisfies the Fermi-Dirac statistics.
In what follows, we fix
$\chi = -1$ with which we can easily compare the pair functions in the
PC and APC and make the subscript $n$ explicit
only when necessary. 

At the vortex core, the pairing symmetry is modified by the synthesis
of the angular momentum. The spatial phase winding around the vortex
is transferred to the pair function \cite{Kato_JPSJ_2000, Hayashi_PRB_2002, Hayashi_JLTP_2003, Tanuma_PRL_2009, Daino_PRB_2012, Ueda_PRB_2021}: 
$f(\hat{\bs{k}})|_{\rho=0} \approx
\bar{f}(\hat{\bs{k}}) e^{i M \phi_k}$, where
$\bar{f}(\hat{\bs{k}})= \Delta_\theta k_z e^{-i\phi_k} /
\Omega_\theta$ 
is the bulk pair function with 
$\Omega_{\theta}=\sqrt{\omega_n^2 + \Delta^2_\theta k_z^2}$ and 
$\Delta_\theta=2\Delta_0\sin\theta_k$. 
In a ($d-i d'$)-wave SC, the pair function at the core is
approximately given by 
\begin{align}
	{f(\bs{r}, \hat{\bs{k}})|_{\rho=0, z\gg\xi_0}} = \left\{
  \begin{array}{cl}
	  \Delta_\theta k_z               / \Omega_\theta & \text{for APC,} \\
	  \Delta_\theta k_z e^{-2i\phi_k} / \Omega_\theta & \text{for PC.} \\
	\end{array} \right.
\label{}
\end{align}
This behavior can be confirmed by the numerical simulations. 
The spatial profiles of $C_l^m$ at the vortex core are shown in
Figs.~\ref{fig:pa}(a,b), where (a) $M=+1$ and (b) $-1$, and $\Gamma=0$. 
Far from the surface ($z \gg \xi_0$), the pair function in each case 
is written by $f \sim C_{1}^{0}Y_{1}^{0} + C_{3}^{0}Y_{3}^{0}$ or 
$f \sim C_{3}^{-2}Y_{3}^{-2} + C_{5}^{-2}Y_{5}^{-2}$ 
\footnote{
The coefficients $C_{1}^{0}$, $C_{3}^{0}$, $C_{3}^{-2}$, and  $C_{5}^{-2}$
correspond to ($p_z$)-, ($d_{3z^2-1}$)-, and ($f_{z(x^2-y^2)}-
if_{xyz}$)-wave components.}.

\begin{table}[tb]
	\caption{Symmetry of Cooper pairs. Only the isotropic Cooper pair
	can survive under the surface roughness and can induce an effective superconductivity. 
 The even- (odd-) parity 
    components are even- (odd-)frequency Cooper pairs.}
	\label{fig:Fp}
	\centering
	\includegraphics[width=1.0\columnwidth]{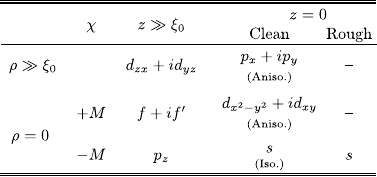}
\end{table}

In addition to the pairing-symmetry conversion at the vortex
core, the surface reflection causes another symmetry conversion. 
The flat-band ABS causes strong inhomogeneity through the suppression
of $\Delta(\bs{r})$ (See the SM \cite{SM} for details). 
The inhomogeneity breaks the local inversion symmetry and triggers the
parity mixing of Cooper pairs\cite{tanaka_PRB_2007, Tanaka_PRL_2007Jul}. The symmetry conversion by the ABSs 
can be approximately expressed as 
$f(\bs{r}, k_\parallel,  k_z ) \to 
 f(\bs{r}, k_\parallel, |k_z|)$ at $z=0$.  
The details of the symmetry conversion by surface ABSs are explained
in the SM \cite{SM}. The pair function at the core-surface
intersection is therefore given by 
\begin{align}
	f(\bs{r}, \hat{\bs{k}})|_{\rho=z=0} = \left\{
  \begin{array}{cl}
	  \Delta_\theta |k_z|               / \Omega_\theta & \text{for APC, } \\
	  \Delta_\theta |k_z| e^{-2i\phi_k} / \Omega_\theta & \text{for PC.} \\
	\end{array} \right.
  \label{}
\end{align}
In the APC, $f|_{\rho=z=0}$ is dominated by the
angle-independent $C_0^0$ (i.e., $s$-wave) component, which can be confirmed 
in Fig.~\ref{fig:pa}(a). 
On the contrary, in the PC, the $C_2^{-2}$ and $C_4^{-2}$ 
components are dominant [Fig.~\ref{fig:pa}(b)] which are anisotropic
in momentum space. 

The isotropic and anisotropic pair functions (e.g., $C_0^0$ and $C_{l \neq 0}^{m}$) show different behaviors against disorder.
The $\Gamma$ dependences of $C_0^0$ and $C_2^{-2}$ are shown in Fig.~\ref{fig:pa}(c). In the APC, the $C_0^0$ component survives even under strong roughness (i.e., $\ell<\xi$). 
the $C_0^0$ component becomes even more robust at low temperatures [Fig.~\ref{fig:pa}(c)]. 
On the contrary, the $C_2^{-2}$ component in the PC is suppressed significantly even with weak roughness ($\ell>\xi$). 

Under the disorder ($\Gamma \neq 0$), the $C_0^0$ component can induce an \textit{effective} $s$-wave pair potential.  
The effective pair potential, the off-diagonal part of $\hat{H}$, is given in  Eq.~\eqref{eq:eff}. 
Importantly, only the isotropic pair function can contribute to $\upeta$ because of the angle average on the Fermi surface (i.e., $\langle f (\bs{k}) \rangle = C_0$). 
When $M=-\chi$, the isotropic $s$-wave Cooper pairs are  induced at the core-surface intersection by the symmetry conversions of Cooper pairs and generate an effective $s$-wave superconductivity. When $M=\chi$, on the contrary, no effective superconductivity is expected because $s$-wave Cooper pair is absent there. 
This is the reason why the robustness of the ZED strongly depends on
$\chi$. 


\textit{Conclusions.}--We have studied the quasiparticle
local density of states (LDOS) of the vortex bound state at the top
surface of a ($d_{zx}+i \chi d_{yz}$)-wave superconductor utilizing
the quasiclassical Eilenberger theory. 
The
LDOS strongly depends on the relation between the chirality $\chi$ and
the vorticity $M$. When $\chi=-M$ (i.e., antiparallel configuration), 
the characteristic dip survives even under significant
roughness. In contrast, the dip structure for $\chi=+M$ (i.e., parallel) is smeared out even under weak roughness. 
By observing this difference by the scanning tunnel spectroscopy (STS)
technique, one can locally detect $\chi(\bs{r})$ and can potenntially make a
chirality-domain map. 
 
The chirality decides the pairing symmetry of subdominant Cooper pairs
induced at the core-surface intersection. 
At the core of \textit{antiparallel} vortex, the $s$-wave Cooper pairs
are induced as the leading term. The isotropic $s$-wave pairs are
robust against disorder and support the characteristic dip structure.
At the \textit{parallel} vortex core, on the other hand, the
($d_{x^2-y^2}-i d_{xy}$)-wave pairs are induced. Such anisotropic
Cooper pairs are fragile against disorder.

\begin{acknowledgments}
	The authors are grateful to V.~Grinenko for the fruitful discussions. 
S.~Y. would like to thank the
``Nagoya University Interdisciplinary Frontier Fellowship'' supported
by JST and Nagoya University.
S.-I.~S. is supported by JSPS Postdoctoral Fellowship for Overseas
Researchers and grateful for the hospitality by the University of
Twente. 
This work was supported by Scientific Research (A)
(KAKENHI Grant No. JP20H00131), and Scientific Research
(B) (KAKENHI Grant No. JP20H01857). 
\end{acknowledgments}




%

%
%
%
%
%
%
%
%
\appendix
\section{Riccati parametrization and boundary condition}
The Eilenberger equation (\buf{1}) can be reduced into two Riccati-type differential equations which provide us stable solutions in the numerical simulations. 
We here introduce the coherence function
$\gamma(\bs{r},\hat{\bs{k}},i\omega_n)$
\cite{Schopohl_PRB_1995, Eschrig_PRB_00, Eschrig_PRB_09} and the boundary
	conditions for $\gamma$ at the specular surface.

The quasiclassical Green's function (GF) can be parametrized in terms of the coherence function $\gamma$, 
\begin{align}
	g&=\frac{1-\gamma\ut{\gamma}}{1+\gamma\ut{\gamma}},\quad
	f=\frac{2\gamma}{1+\gamma\ut{\gamma}}.\label{coh_fun}
\end{align}
Substituting Eq.~\eqref{coh_fun} into Eq.~(\buf{1}) in the main text, 
we find that the coherence functions obey the Riccati-type differential equations, 
\begin{align}
	\bs{v}_F\cdot \bs{\nabla} \gamma-
	2\left(\omega_n +\upxi \right)
	\gamma-\upeta
	+\upeta^*\gamma^2
	=0,\label{riccati}
	\\
	\bs{v}_F\cdot \bs{\nabla} \undertilde{\gamma}+
	2\left(\omega_n+\upxi \right)
	\undertilde{\gamma}
	+\upeta^*
	-\upeta \undertilde{\gamma}^2
	=0.
	\label{undertilde_riccati}
\end{align}
We can obtain the GF by solving Eq.~\eqref{riccati} and \eqref{undertilde_riccati} along the quasiclassical paths. 
The coherence functions in the homogeneous limit (i.e., $\bs{\nabla} \gamma =0$) are given as
\begin{align}
	& \bar{\gamma}
	=
	\frac{\Delta_\infty}
	{\omega_n+\sqrt{\omega_n^2+|\Delta_\infty|^2}},
  \\
	& \bar{\ut{\gamma}}
	=
	\frac{\Delta_\infty^*}
	{\omega_n+\sqrt{\omega_n^2+|\Delta_\infty|^2}},
	\label{coh_fun_bulk2}
\end{align}
where 
$\Delta_\infty = \Delta_\infty (\hat{\bs{k}})$ is the
momentum-dependent pair potential in a
uniform system and $\Gamma=0$.

We assume the surface at $z=0$ is specular.
Therefore, the boundary conditions for the coherence function $\gamma$ can be written as, 
\begin{align}
  & \gamma
	(\bs{r}_{\parallel}, 
	z=0, 
  \hat{\bs{k}}_{\parallel}, 
  k_\perp, 
  i\omega_n)
	\\
  & =
	\gamma
	(\bs{r}_{\parallel}, 
	z=0, \hat{\bs{k}}_{\parallel},
  -k_\perp, 
  i\omega_n),
  \label{bc}
\end{align}
where $\hat{\bs{k}}_\parallel=(k_x,k_y)$ and $k_\perp=k_z$.

\section{Spherical harmonics}
We use the complex spherical harmonics $Y_l^{m}$ to expand the pair amplitude.
The complex spherical harmonics are normalized to satisfy the orthonormal relation:
\begin{align}
	\int Y^{m}_l Y^{m^\prime*}_{l^\prime}d\Omega&=\delta_{ll^\prime}\delta_{mm^\prime}.
\end{align}
with $\delta_{ll'}$ being the Kronecker delta.  
We summarize the analytical expressions of $Y_l^{m}$ in Table.~\ref{table:Yae}.

\begin{table}[tb]
	\caption{Analytical expression of $Y^{m}_l$}
  \label{table:Yae}
  \centering
  \begin{tabular}{ccc}
    \hline
    ~ $l$ & ~ ~$m$~ ~ & Complex spherical harmonics $Y^{m}_l$ \\
    \hline \hline
    $0$ & $ 0$ & $ \sqrt{({1}  /{ 4\pi})}$                                           \\[3mm]
    $1$ & $-1$ & $ \sqrt{({3}  /{ 8\pi})}e^{-i\phi}\sin\theta$                       \\[1mm]
    ~~~ & $ 0$ & $ \sqrt{({3}  /{ 4\pi})}\cos\theta$                                 \\[1mm]
    ~~~ & $ 1$ & $-\sqrt{({3}  /{ 8\pi})}e^{i\phi}\sin\theta$                        \\[3mm]
    $2$ & $-2$ & $ \sqrt{({15} /{32\pi})}e^{-2i\phi}\sin^2\theta$                    \\[1mm]
    ~~~ & $-1$ & $ \sqrt{({15} /{ 8\pi})}e^{-i\phi}\sin\theta\cos\theta$             \\[1mm]
    ~~~ & $ 0$ & $ \sqrt{({5}  /{16\pi})}\left(3\cos^2\theta-1\right)$               \\[1mm]
    ~~~ & $ 1$ & $-\sqrt{({15} /{ 8\pi})}e^{i\phi}\sin\theta\cos\theta$              \\[1mm]
    ~~~ & $ 2$ & $ \sqrt{({15} /{32\pi})}e^{2i\phi}\sin^2\theta$                     \\[3mm]
    $3$ & $-3$ & $ \sqrt{({35} /{64\pi})}e^{-3i\phi}\sin^3\theta$                    \\[1mm]
    ~~~ & $-2$ & $ \sqrt{({105}/{32\pi})}e^{-2i\phi}\sin^2\theta\cos\theta$          \\[1mm]
    ~~~ & $-1$ & $ \sqrt{({21} /{64\pi})}e^{-i\phi}\sin\theta\left(5\cos^2-1\right)$ \\[1mm]
    ~~~ & $ 0$ & $ \sqrt{({7}  /{16\pi})}\left(5\cos^3\theta-3\cos\theta\right)$     \\[1mm]
    ~~~ & $ 1$ & $-\sqrt{({21} /{64\pi})}e^{i\phi}\sin\theta\left(5\cos^2-1\right)$  \\[1mm]
    ~~~ & $ 2$ & $ \sqrt{({105}/{32\pi})}e^{2i\phi}\sin^2\theta\cos\theta$           \\[1mm]
    ~~~ & $ 3$ & $-\sqrt{({35} /{64\pi})}e^{3i\phi}\sin^3\theta$                     \\[1mm]
    \hline
  \end{tabular}
\end{table}

\section{Symmetry of pair amplitudes at the core-surface intersection}\label{sec:analytic}

In this section, we discuss the symmetry of pair amplitudes at the
core-surface intersection focusing on the symmetry and boundary
condition of the system.  Assuming the uniform pair potential, we can
obtain the analytic expression of the pair amplitude at the
core-surface intersection.


We assume the rotational symmetry around the $z$ axis: the
pair potential in Eq.~(\buf{4}) is changed only the phase by the
rotation operation
$(\phi_k,\Phi) \to (\phi_k+\varphi,\Phi+\varphi)$ as
$\Delta \to \Delta e^{i(\chi+M)\varphi}$.
Under the rotation operation, the Eilenberger equation (\buf{1}) is rewritten as
\begin{align}
   &\bs{v}_F \cdot \bs{\nabla} \check{g}_\varphi
   +[\tau_3 \omega_n + \check{R}_\varphi\check{H}\check{R}^\dagger_\varphi,\check{g}_\varphi]=0,\\
   &\check{R}_\varphi=\left( \begin{array}{cc}
        e^{i(\chi+M)\varphi/2} & 0  \\
                             0 & e^{-i(\chi+M)\varphi/2}
       \end{array}
   \right),
\end{align}
where
$\check{g}_\varphi =\check{g}(\rho,z,\Phi+\varphi,\theta_k,\phi_k+\varphi,i\omega_n)$
and we set $\Gamma=0$ for simplicity.
We find that
$\check{R}^\dagger_\varphi\check{g}_\varphi\check{R}_\varphi$ is also
the solution of Eq.~(\buf{1}){: $\check{g}=\check{R}^\dagger_\varphi\check{g}_\varphi\check{R}_\varphi$}.
Setting $\varphi=-\phi_k$, we obtain
\begin{align}
 &\check{g}(\rho,z,\Phi,\theta_k,\phi_k,i\omega_n)
 \notag \\
 &=
 \check{R}^\dagger_{-\phi_k}
 \check{g}(\rho,z,\Phi-\phi_k,\theta_k,0,i\omega_n)
 \check{R}_{-\phi_k}.
\end{align}
The pair amplitude at the off-diagonal parts satisfies
\begin{align}
  & f(\rho,z,\Phi,\theta_k,\phi_k,i\omega_n)
	\notag \\
  &=f(\rho,z,\Phi-\phi_k,\theta_k,0,i\omega_n)e^{i(\chi+M)\phi_k}.
\end{align}
The pair amplitude $f$ at $\rho=0$ should not depend on $\Phi$ because $f$ must be a 
single-valued function. Therefore, $f$ is given by 
\begin{align}
	f(\rho=0,z,\Phi, \bs{k},
	i\omega_n)=F(z,\theta_k,i\omega_n)e^{i(\chi+M)\phi_k},\label{f_bc1}
\end{align}
where we define $F(z,\theta_k,i\omega_n)=f(\rho=0,z,\Phi,\theta_k,\phi_k=0,i\omega_n)$.
From the boundary condition~\eqref{bc}, $F$ satisfies
\begin{align}
	F(z=0,\theta_k,i\omega_n)=F(z=0,\pi-\theta_k,i\omega_n).\label{f_bc2}
\end{align}
Combining Eqs.~\eqref{f_bc1} and \eqref{f_bc2}, one finds that $f$ at the core-surface intersection 
must be even-parity when $\chi+M$ is an even integer. 
Namely, only the even-frequency spin-singlet even-parity components
are allowed at the core-surface intersection.

The quasiclassical GF in spin-singlet SCs satisfy $g=
(1-f\undertilde{f})^{1/2}$ derived from the normalization condition
($\check{g}^2=\check{1}$) in Eq.~(\buf{1}).  When $f$ is an even-function of $\omega_n$, this
condition is reduced to $g=\sqrt{1-|f|^2}$.  The existence of
even-frequency pairs suppresses the local density of states (LDOS)
[i.e., $N(E) \sim \langle\mathrm{Re}[g]\rangle$] as happened in the homogeneous SC.

The robustness of the locally induced even-frequency pairs can be
examined by focusing on their momentum dependence.  For simplicity, we
assume that the pair potential is uniform and that the vortex
core is infinitely small:
$\Delta=\Delta_0\psi_\chi(\hat{\bs{k}})e^{iM\Phi}$\cite{Thuneberg_PRB_1984,
Hayashi_PRB_2002}.  
The zero-energy pair amplitude at the core-surface intersection can be
approximately calculated as
\begin{align}
	F(z=0,\theta_k,i\delta)
	=\frac{\Delta_0|\sin2\theta_k|}{\sqrt{\delta^2+\Delta_0^2\sin^22\theta_k}}. \label{f_sym}
\end{align}
The function $F$ satisfies $F(z=0,\theta_k,i\delta) \approx 1$ when $\sin 2\theta_k \neq 0$.  
Substituting Eq.~\eqref{f_sym} into Eq.~\eqref{f_bc1}, we find that the
isotropic $s$-wave pairs are induced when $\chi=-M$ due to the
cancellation of the chirality and the vorticity.  In the parallel
configuration (PC), $f$ has the phase factor $e^{2iM\phi_k}$.  
The isotropic s-wave Cooper pair is expected to be robust against the
non-magnetic impurities because the pair amplitude does not depend on
the momentum.  The appearance of the isotropic $s$-wave Cooper pair
results in a robust dip structure at the zero energy.  On the other
hand, the zero-energy dip is fragile when $\chi=M$ due to the
momentum-dependent phase winding $e^{2iM\phi_k}$. The
random scatterings by non-magnetic impurities disturb the constructing
interference of Cooper pairs. 

\begin{figure*}[tb]
	\centering
	\includegraphics[width=0.98\textwidth]{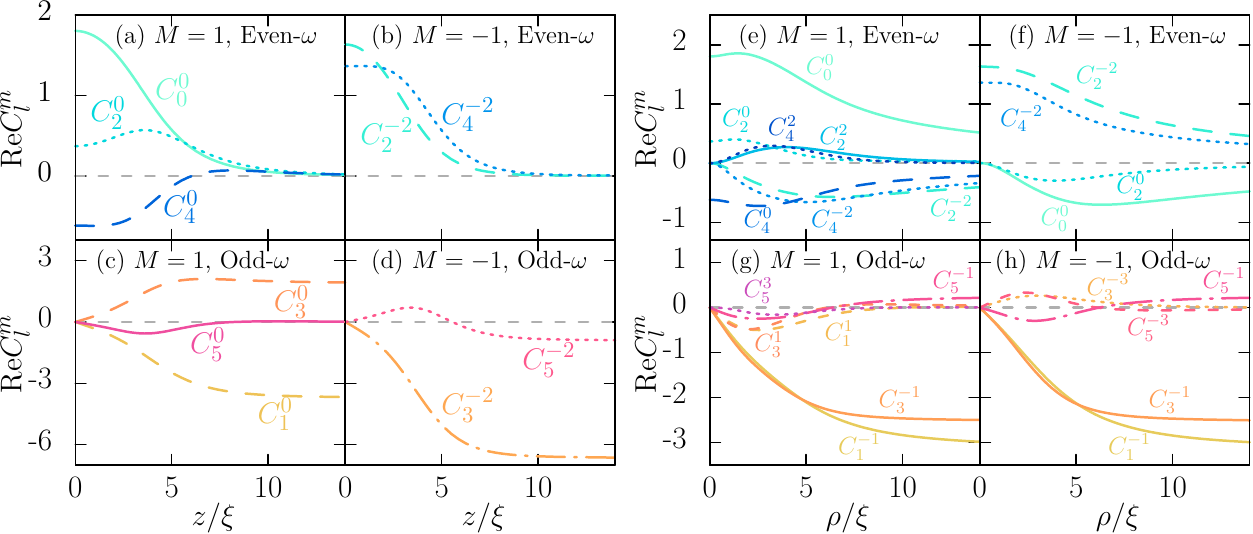}
	\caption{
		Pair amplitudes at (a-d) the vortex core
		and (e-h) the clean surface. 
		The results for the APC $M=1$ and PC $M=-1$ are respectively shown in (a,c,e,g) and (b,d,f,h).  
		The even-frequency (odd-frequency) components are shown in the upper (lower) panels.  
		The imaginary part of the pair amplitudes is
negligibly small. }
	\label{fig:pa_sm}
\end{figure*}

\begin{figure*}[tbp]
	\centering
	\includegraphics[width=0.98\textwidth]{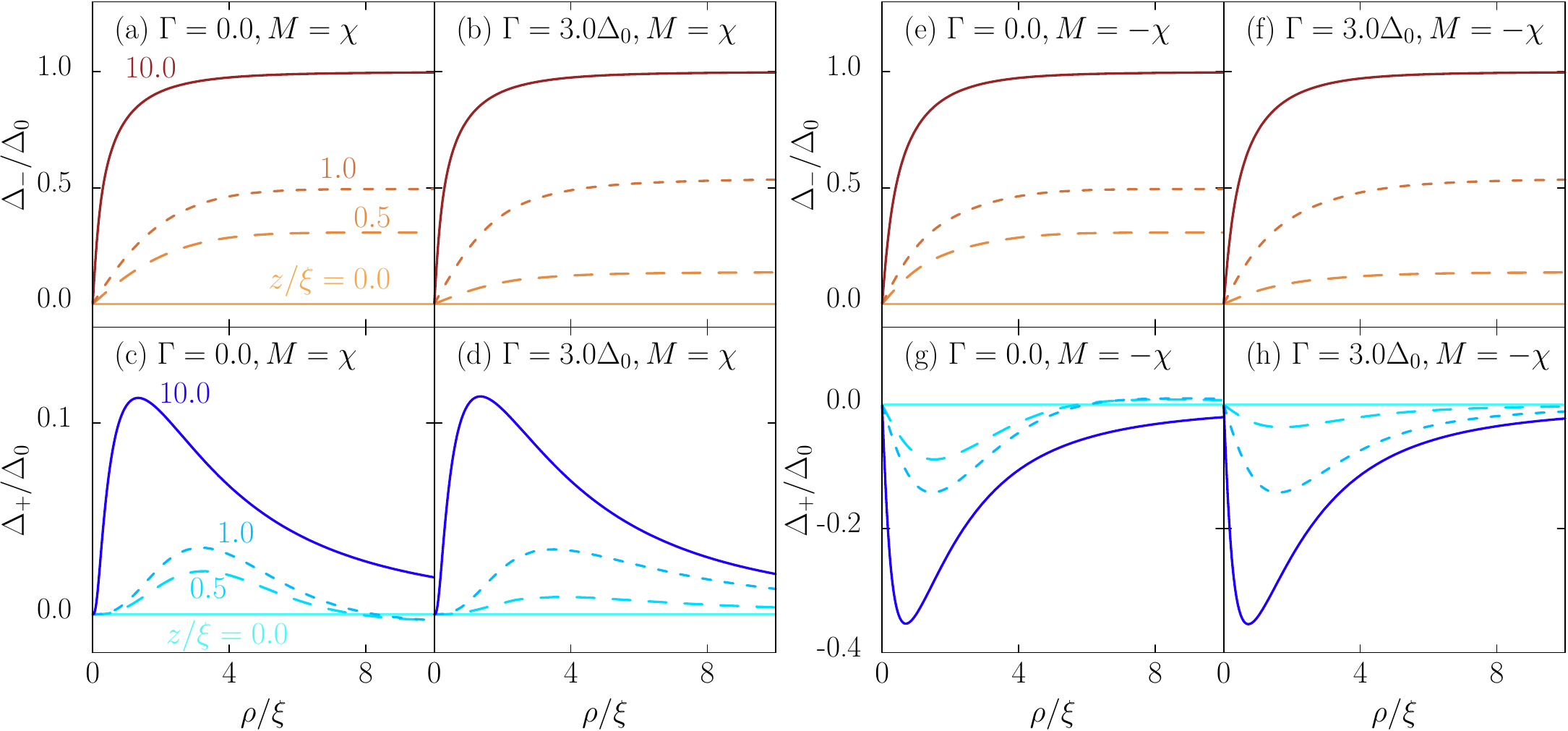}
	\caption{
		Spatial dependences of the dominant and subdominant components ($\Delta_-$ and $\Delta_+$) of the pair potential.
		The dominant component $\Delta_-$ is set to a positive value.
		The pair potential is normalized by its bulk value $\Delta_0$.
		The depth is set to $z/\xi=0.0$, $0.5$, $1.0$, $2.0$, $4.0$ and $10.0$.
		The impurity-scattering rate is set to $\Gamma/\Delta_0=0.0$ and $3.0$.
	}
	\label{sup:fig:pf}
\end{figure*}
\begin{figure*}[tbp]
	\centering
	\includegraphics[width=0.98\textwidth]{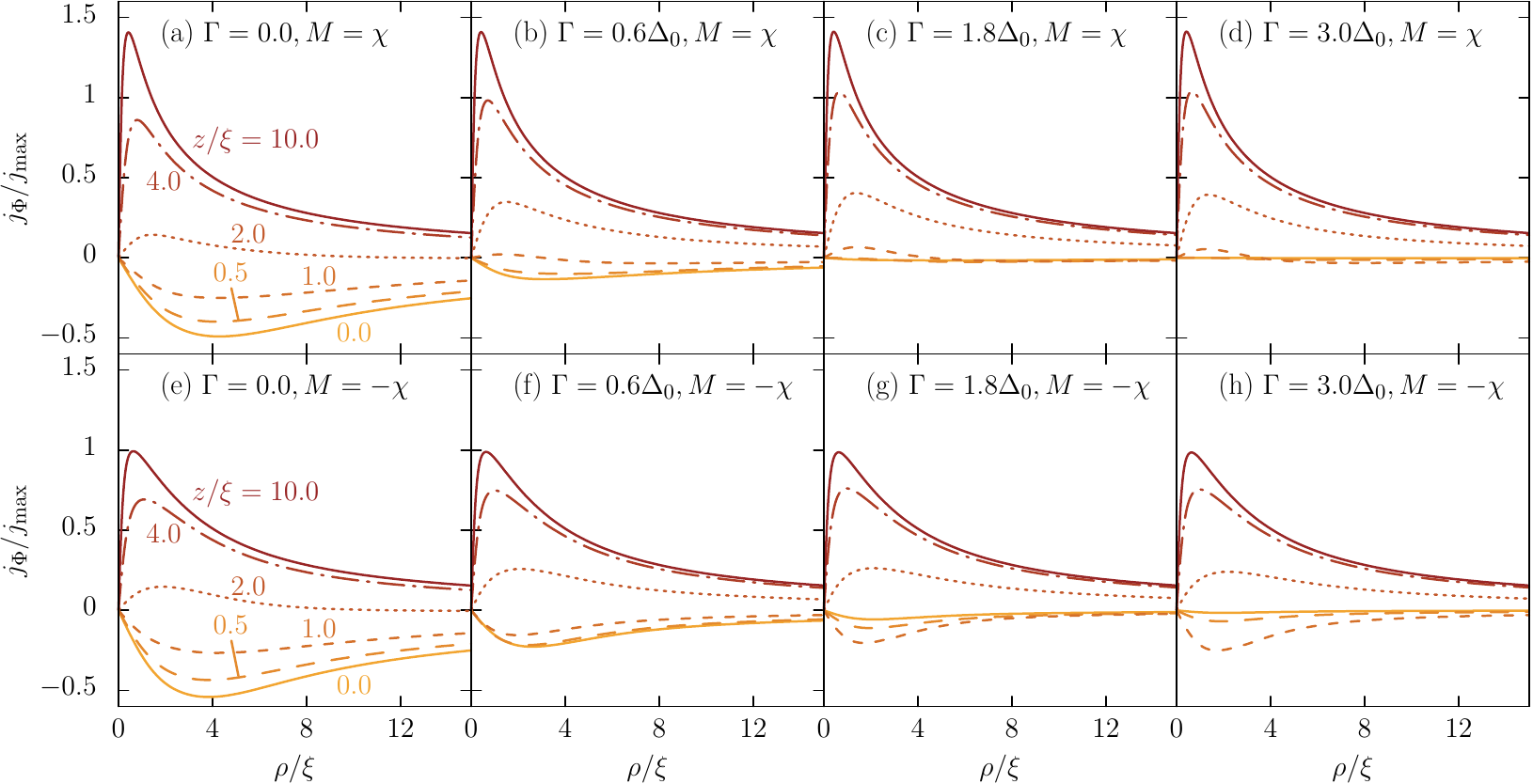}
	\caption{Spatial profiles of current density in the presence of surface roughness.
		The current density is normalized by $j_{\mathrm{max}}$ being the maximum value of $j_\Phi$ at $T/T_c=0.2$ and $M=-\chi$.
		The depth is set to $z/\xi=0.0$, $0.5$, $1.0$, $2.0$, $4.0$ and $10.0$.
		The impurity scattering rate is set to $\Gamma/\Delta_0=0.0$,
		$0.6$, $1.8$ and $3.0$.
	}
	\label{sup:fig:j}
\end{figure*}

\section{Numerical results of pair amplitudes}

The pair amplitudes at the vortex core and at the surface 
are respectively shown in
Fig.~\ref{fig:pa_sm}(a-d) and \ref{fig:pa_sm}(e-h).
The even- and odd-frequency components are shown in Fig.~\ref{fig:pa_sm}(a,b,e,f) and \ref{fig:pa_sm}(c,d,g,h) respectively.
We do not show the components whose maximum value is less than
$C_\mathrm{max}/10$, where $C_\mathrm{max}$ is the largest value
of $|C_l^m(\rho,z)|$.

As shown in Figs.~\ref{fig:pa_sm}(a-d), at $z \gg \xi$ and $\rho=0$, the
odd-frequency components are dominant. The local symmetry breaking
induces odd-frequency Cooper pairs at the core \cite{Tanuma_PRL_2009}. 
With approaching the surface, the amplitudes of
the even-frequency components increase [Fig.~\ref{fig:pa_sm}(a,b)], 
whereas the odd-frequency
components vanish at the core-surface intersection [Fig.~\ref{fig:pa_sm}(c,d)]. 
This behavior is consistent with the discussion in Sec.~\ref{sec:analytic}.

The pair amplitudes at the surface are plotted as functions of $\rho$ 
in Fig.~\ref{fig:pa_sm}(e-h).  Far from the vortex core ($\rho \gg \xi$), 
the odd-frequency Cooper pairs are dominant at the surface as shown in
Figs.~\ref{fig:pa_sm}(g,h).  
When flat-band ABSs appear, the pair
potential is significantly suppressed at the surface. This spatial
inhomogeneity induces odd-frequency Cooper pairs \cite{tanaka_PRB_2007}.
At the core-surface intersection, the odd-frequency Cooper pairs disappear, whereas even-frequency components have large amplitudes.

\section{Pair potential}

In this section, we discuss the spatial dependence of the
self-consistent pair potential.  We set $\chi=-1$ in this section:
$\Delta_-$ and $\Delta_+$ are dominant and subdominant components
The dominant and subdominant components of the pair potential are
shown in Fig.~\ref{sup:fig:pf}, where the results for the PC are shown 
in Fig.~\ref{sup:fig:pf}(a-d), those for the antiparallel
configuration (APC) are shown in Fig.~\ref{sup:fig:pf}(e-h), and the phase of $\Delta_-$ at $\rho\gg\xi$ is set to $0$.
The surface roughness is set to
$\Gamma=0$ in Fig.~\ref{sup:fig:pf}(a,c,e,g), and $\Gamma=3.0\Delta_0$ in Fig.~\ref{sup:fig:pf}(b,d,f,h). 
The phase difference between $\Delta_-$ and
$\Delta_+$ is $0$ ($\pi$) in the PC (APC)
which is consistent with Ref.~\cite{Kato_JPSJ_2001}. 
The pair potentials vanish at the vortex core so that the pair
potentials have to be single-valued functions.  The
subdominant component is induced around the vortex core and decays to
zero far from the vortex core ($\rho \gg \xi$). The pair potential is
suppressed near the surface.
The $(d+id')$-wave components are prohibited at the surface due to the boundary
condition \eqref{bc}.

The pair potential is suppressed in the disordered regions
($z=0.5\xi$) regardless of the vorticity. However, the amplitude of 
the pair potential at $z/\xi=0.5$ remains finite even with the strong 
surface roughness ($\Gamma/\Delta_0=3.0$) because the
disorder exists only near the surface.

\section{Current density}

The current density $\bs{j}(\bs{r})$ are given by
\begin{align}
	&\bs{j}(\bs{r})
	=-4|e|N_0\pi k_BT\sum_{n=0}^{N_{\mathrm{c}}}
	\int\frac{d\Omega}{4\pi}
	\bs{v}_F \mathrm{Im}[g(\hat{\bs{k}},\bs{r},i\omega_n)].
	\label{jeq}
\end{align}
In this section, we set the vorticity as $M=-1$ so that the
azimuthal components of the current density $j_\Phi$ is positive deep inside the
SC. We show the current density for
$\Gamma/\Delta_0=0.0$, $0.6$, $1.8$, $3.0$ in Fig.~\ref{sup:fig:j}.  We
normalized the current density by the maximum value for $\chi=-M$.  The
maximum value of $j_\Phi$ for $\chi=M$ is larger than the value
for $\chi=-M$ because the direction of the angular momentum in the
momentum space is the same as the direction of the spatial phase
winding.  The directions of the vortex currents are inverted near the
surface ($z/\xi<1.5$) in the clean limit as shown in
Fig.~\ref{sup:fig:j}(a) and \ref{sup:fig:j}(e) because of the
flat-band ABSs \cite{Suzuki_PRB_2014,
Suzuki_PRB_2015, Yoshida_PRR_22}.

When $\chi=M$, the vortex current in the disordered region
($z/\xi<1.0$) is suppressed in the presence of surface roughness and
disappears at $\Gamma=1.8\Delta_0$ as shown in
Fig.~\ref{sup:fig:j}(b), \ref{sup:fig:j}(c) and \ref{sup:fig:j}(d).
On the other hand, the inverted current remains finite even
with $\Gamma=3.0\Delta_0$ in the APC as shown in
Fig.~\ref{sup:fig:j}(f), \ref{sup:fig:j}(g) and \ref{sup:fig:j}(h).
The depth $z$ at which the current density takes a maximum value is moved to the interface between the disordered and clean regions from the surface with increasing the impurity scattering rate.

\bibliographystyle{apsrev4-1}
\bibliography{Yoshida_no_bib}

\end{document}